\documentclass[twoside]{book}
\usepackage{graphicx}
 \makeatletter
\ifx\UNDEF\mail\def\mail{ }\else\fi
\ifx\UNDEF\prange\def\prange{0 0}\else\fi

\gdef\@empty{}
\def\Mail#1 #2 {\gdef\thecontact{#1}\gdef\theaddr{#2}}
\def\Range#1 #2 {\gdef\thefirstpage{#1}\gdef\thelastpage{#2}}
{\let\'\mail \expandafter\Mail\' }	% do not remove space between ' and }
{\let\'\prange \expandafter\Range\' }	% do not remove space between ' and }
 \gdef\@shtitle{\relax}
 \long\def\shtitle#1{\gdef\@shtitle{#1}}
 \long\def\author#1{\gdef\@author{#1}}
 \def\affil#1{\par\noindent{\rm#1\par}}
 \gdef\@abstract{}
 \long\def\abstract#1{\gdef\@abstract{#1}}

 \renewcommand{\@evenhead}{\thepage\qquad\qquad\@shtitle\hfil}
 \renewcommand{\@oddhead}{\hfil\@shtitle\qquad\qquad\thepage}

 \def\maketitle{\thispagestyle{empty}\chapter{\@title}}
 \renewcommand\chapter{\if@openright\cleardoublepage\else\clearpage\fi
                    \thispagestyle{empty}%
                    \global\@topnum\z@
                    \@afterindentfalse
                    \secdef\@chapter\@schapter}
 \def\@makechapterhead#1{%
  \vspace*{50\p@}%
  {\parindent \z@ \raggedleft \normalfont
    \ifnum \c@secnumdepth >\m@ne
      \if@mainmatter
        \huge \@chapapp{} \thechapter
        \par\nobreak
        \vskip 20\p@
      \fi
    \fi
    \interlinepenalty\@M
    \Huge \bfseries #1\par\nobreak
    \vskip.25in
    \large\bfseries\@author\par\nobreak
    \vskip 40\p@}
    \ifx\@abstract\@empty\else{\small\@abstract\par\vskip20\p@}\fi
  }

\DeclareRobustCommand\em
        {\@nomath\em \ifdim \fontdimen\@ne\font >\z@
                       \upshape \else \slshape \fi}

\def\@begintheorem#1#2{\sl \trivlist \item[\hskip \labelsep{\bf #1\ #2}]}
\def\@opargbegintheorem#1#2#3{\sl \trivlist
     \item[\hskip \labelsep{\bf #1\ #2\ (#3)}]}

  \def\@arabic#1{\number #1} % my redefinition

\long\def\@makecaption#1#2{
	\vskip\abovecaptionskip
	\sbox\@tempboxa{{\small {\bf #1}: #2}}%
	\ifdim\wd\@tempboxa>\hsize
	    {\small {\bf #1}: #2\par}
	\else
	   \global\@minipagefalse
	   \hbox to\hsize{\hfil\box\@tempboxa\hfil}
	\fi
	\vskip \belowcaptionskip}

\def\figstrut#1{\hbox to\linewidth{\vrule height#1\hfill}}

\makeatother

\shtitle{Protocol Requirements for Self-organizing Artifacts}
\title{Protocol Requirements for Self-organizing Artifacts: Towards 
an Ambient Intelligence}
\author{Carlos Gershenson and Francis Heylighen \affil{Centrum Leo 
Apostel, Vrije Universiteit Brussel, Belgium \\ 
\{cgershen,fheyligh\}@vub.ac.be}}
\abstract{We discuss which properties common-use artifacts should 
have to collaborate without human intervention. We conceive how 
devices, such as mobile phones, PDAs, and home appliances, could be 
seamlessly integrated to provide an ``ambient intelligence" that 
responds to the user's desires without requiring explicit programming 
or commands. While the hardware and software technology to build such 
systems already exists, as yet there is no standard protocol that can 
learn new meanings. We propose the first steps in the development of 
such a protocol, which would need to be adaptive, extensible, and 
open to the community, while promoting self-organization. We argue 
that devices, interacting through ``game-like'' moves, can learn to 
agree about how to communicate, with whom to cooperate, and how to 
delegate and coordinate specialized tasks. Thus, they may evolve a 
distributed cognition or collective intelligence capable of tackling 
complex tasks.}

\begin{document}
\maketitle

\section{A Scenario}

The diversity and capabilities of devices we use at home, school, or 
work, are increasing constantly. The functions of different devices 
often overlap (e.g. a portable computer and a mobile phone have 
agendas; a radio-clock and a PDA have alarms), but most often we 
cannot combine their capabilities automatically (e.g. the PDA cannot 
tell the radio to set its alarm for the early Tuesday's appointment), 
and users need to repeat the same tasks for different devices (e.g. 
setting up an address book in different devices). Moreover, using the 
functionality of some devices in combination with others would be 
convenient (e.g. if my computer has an Intelligent User Interface, I 
would like to use it to ask for coffee, without the need of having 
speech recognition in the coffee machine: The computer should be able 
to ask the coffee machine for  cappuccino).

	Could we build devices so that they would \emph{automatically 
coordinate}, combining their functions, and possibly producing new, 
``emergent" ones? The technology to achieve this is already at hand. 
What we lack is a proper design methodology, able to tackle the 
problems posed by autonomously communicating artifacts in a 
constantly changing technosphere. In this paper we try to delineate 
the requirements that such a design paradigm should fulfill. The 
scenario we imagine considers a nearby future where technological 
artifacts \emph{self-organize}, in the sense that they are able to 
communicate and perform desirable tasks with minimal human 
intervention.

	This vision is closely related to the concept of ``Ambient 
Intelligence" (AmI)\cite{ISTAG2001}, which envisages a future where 
people are surrounded by ``smart" and ``sensitive" devices. AmI would 
be the result of the integration of three technologies: Ubiquitous 
Computing \cite{Weiser1993}, Ubiquitous Communication, and 
Intelligent User Friendly Interfaces. The first one conceives of a 
seamless integration of computation processes taking place in the 
variety of artifacts that surround us, being part of ``The Grid", the 
network that would allow anyone anywhere to access the required 
computing power. The present paper focuses on the aspect of 
Ubiquitous Communication that attempts to obtain seamless information 
exchange between devices. Intelligent User Friendly Interfaces should 
enable an intuitive, effortless interaction between users and devices.

	With current approaches, this scenario would be possible, 
since we have the technology, but extremely expensive, since people 
would need to buy from the same producer all of their devices. We can 
see a similar casein the area of Home Automation: the technology is 
available on the market, but it is not possible to buy today 
ventilation for a house, and in five years integrate the system with 
a new fire detector. An engineer needs to integrate them manually, so 
that the ventilation system could be activated if smoke is detected, 
simply because the ventilation system was not designed to receive 
such signals. These limitations increase the price and restrict the 
market of devices for Home Automation, since complete solutions 
should be bought in order to have full coordination and functionality 
between devices. People would be more willing to invest in Home 
Automation if they could have the possibility of acquiring it 
progressively.

\section{Requirements for self-organizing artifacts}

	We see self-organization as a paradigm for designing, 
controlling, and understanding systems 
\cite{GershensonHeylighen2003a}. A key characteristic of a 
self-organizing system is that structure and function of the system 
``emerge" from interactions between the elements. The purpose should 
not be explicitly designed, programmed, or controlled. The components 
should \emph{interact} freely with each other and with the 
environment, mutually adapting to reach an intrinsically 
``preferable" or ``fit" configuration (attractor), thus defining an 
emergent purpose for the system \cite{HeylighenGershenson2003}. By 
``self-organizing artifacts" we mean a setup where different devices, 
with different fabrications and functionalities, and moving in and 
out of different configurations, can communicate and integrate 
information to produce novel functionalities that the devices by 
themselves could not achieve.

	A first requirement for such communication is cross-platform 
compatibility. This is already achieved for programming with Java, 
and for documents with XML. Another requirement is wireless 
communication, which is offered by technologies such as IR, Bluetooth 
and WiFi. Near Field Communications (NFC) is a newly envisioned 
standard, proposed by a consortium headed by Sony, Nokia, and 
Philips, which would allow information to be transmitted between 
devices that come in close spatial proximity (``touching").

	Even with such a standard, the problem remains that the user 
generally would need to specifically request such communication 
between devices (e.g. ``transfer this file from here to there"). 
Ideally, the devices would \emph{know} what we want them to do and 
how to do it. User Interfaces already help us to tell them our 
wishes. Still, one device cannot tell \emph{another} device what we 
want, especially if they are produced by different manufacturers. 
This is a general problem of communication between artifacts: they 
can recognize standard messages, but they do not ``know" what the 
messages \emph{mean}. To avoid endless debates, we can say that the 
meaning of a message is determined by its \emph{use} 
\cite{Wittgenstein1999}: if a device has received a message, and does 
``the right thing" (for the user), then it has ``understood" the 
meaning of the message. Thus, the user's satisfaction is the ultimate 
measure of the effectiveness of the artifacts' performance.

	Another issue is how to deal with changes in technology. We 
do not want to reconfigure every artifact each time a new device 
arrives. Moreover, we want the old devices to be able at least to 
cope with the functionality of new ones. New devices should configure 
themselves as automatically as possible. Older ones may require user 
intervention at first (as they cannot know beforehand which functions 
will be required), but they should be able to cope with new 
technology being added to the network. The overall system must be 
\emph{adaptive}, \emph{extensible}, and \emph{open}.

	An adaptive system can cope with unexpected changes in its 
environment, as exemplified by the constantly changing technology. 
Having flexibility built into our systems is desirable: they should 
at least be able to tolerate events they were not designed for 
without breaking down, but preferably try to find adapted solutions, 
or at least ask assistance from the user. For example, home 
appliances have a limited set of functions. To have them 
self-organize (e.g. the alarm clock coordinating with the microwave 
oven, and the oven with the kettle), their functions could be easily 
programmed to respond to unknown messages. If a new device arrives, 
and an old one does not know what to do when it receives a message, 
it can check what the user wants, thus learning how to respond 
appropriately. The possibility to add more devices to an existing 
configuration may be called \emph{extensibility}.

	Suppose that a company develops adaptable and extensible 
devices that interact seamlessly with each other. This would still 
leave the problem that customers cannot add devices from other 
companies, as these would follow their own standards, thus creating 
compatibility problems. We believe that the solution is to have 
\emph{open} technologies, in the spirit of GNU. Open means that 
everyone has free access to their specifications. The advantage is 
that they can develop much faster, meeting the requirements of more 
people, because they are developed by a global community that can try 
out many more approaches than any single company. Still, a company 
can benefit in promoting an open technology, since this would provide 
them with free publicity while everyone is using their protocol (e.g. 
Sun's Java).

	Open technology can respond to the needs of the largest 
variety of people, while allowing problems and errors to be detected 
and corrected more easily. Another advantage is that it allows people 
to get updates developed by other users for free. For example, if I 
program my ``old" toaster to integrate with my new mobile phone, it 
costs me nothing to make the program available on the Internet to 
anyone else who might need it. Thus, updates, extensions, and 
specialized applications can flow much more quickly from a global 
community than from a private company. Still, companies would benefit 
from this approach, since people would be more willing to buy new 
devices as integrating them into their existing, open setup, will be 
easier.

\section{Achieving self-organization}

We can divide the problem of self-organizing integration into three 
subproblems: 1) devices should learn to \emph{communicate} with each 
other, even when they have no a priori shared understanding of what a 
particular message or function means; 2) devices should learn which 
other devices they can trust to \emph{cooperate}, avoiding the 
others; 3) devices should develop an efficient \emph{division of 
labour} and workflow, so that each performs that part of the overall 
task that it is most competent at, at the right moment, while 
delegating the remaining functions to the others.

	These issues are all part of \emph{collective intelligence} 
\cite{Heylighen1999} or \emph{distributed cognition} 
\cite{Hutchins1995}: a complex problem cannot be tackled by a single 
device or agent, but must be solved by them working together, in an 
efficiently coordinated, yet spatially distributed, system, where 
information flows from the one agent to the other according to 
well-adapted rules. Until now, distributed cognition has been studied 
mostly in existing systems, such as human organizations 
\cite{Hutchins1995} or animal ``swarms" \cite{BonabeauEtAl1998}, that 
have evolved over many generations to develop workable rules. Having 
the rules self-organize from scratch is a much bigger challenge, 
which has been addressed to some degree in distributed AI and 
multi-agent simulations of social systems. Inspired by these first 
explorations, we will propose a number of general mechanisms that 
could probably tackle the three subproblems. However, extensive 
simulation will clearly be needed to test and elaborate these 
mechanisms.

\section{Learning to communicate}

To communicate effectively, different agents must use the same 
concepts or categories. To achieve effective coordination, agents 
must reach a shared understanding of a concept, so that they agree 
about which situations and actions belong to that category, and which 
do not. A group of agents negotiating such a consensus may 
self-organize, so that a globally shared categorisation emerges out 
of local interactions between agents.

	Such self-organization has been shown in different 
simulations of the evolution of language 
\cite{Steels1998,HutchinsHazelhurst1995}. Here, interacting software 
agents or robots try to develop a shared lexicon, so that they 
interpret the same expressions, symbols, or ``words" in the same way. 
In these simulations agents interact according to a protocol called a 
``language \emph{game}". There are many varieties of such games, but 
the general principle is that two agents ``meet" in virtual space, 
which means that through their sensors they experience the same 
situation at the same time. Then they try to achieve a consensus on 
how to designate one of the components of their shared experience by 
each in turn performing elementary \emph{moves}.

	In a typical move, the first agent produces an ``utterance" 
referring to a phenomenon that belongs to one of its inbuilt or 
previously learned categories, and the second one finds the best 
fitting category for that phenomenon in its knowledge base. The 
second agent then indicates a phenomenon belonging to that same 
category. If this phenomenon also belongs to the same category for 
the first agent, both categorisations are reinforced, otherwise they 
are reduced in strength. In the next move of the ``game", another 
phenomenon is indicated, which may or may not belong to the category. 
The corresponding categorisation is strengthened or weakened 
depending on the degree of agreement. After a number of moves the 
game is stopped, each agent maintaining the mutually adjusted 
categories. Each agent in turn is coupled to another agent in the 
system, to play a new game using different phenomena. After some 
games a stable and coherent system of categories shared by all agents 
is likely to emerge through self-organization. A good example of such 
a set-up can be found in Belpaeme's \cite{Belpaeme2001}  simulation 
of the origin of shared colour categories.

	If for some reason devices are not able to communicate, they 
should be able to notify the user, and ask for the correct 
interpretation of the message. This is easy, since devices have a 
limited functionality. It would be possible to ``teach" a device what 
to do if it receives a particular message, and the device should 
``learn" the meaning of the message.

	Research has been done in multi-agent systems where agents 
negotiate their protocols \cite{ReedEtAl2001, DastaniEtAl2001}, which 
could be extended for a setup of self-organizing artifacts. However, 
agent communication standards, such as FIPA, still do not contemplate 
adaptation to new meanings. Nevertheless, there is promising research 
going on in this direction.

\section{Learning to cooperate}

Integrated devices should not only communicate, but cooperate. 
Cooperation may seem self-evident in preprogramed systems, where the 
components are explicitly designed to respond appropriately to 
requests made by other components. However, this is no longer the 
case in open, extensible configurations.

	For example, a person at the airport would like her PDA to 
collaborate with the devices present at the airport, so that it can 
automatically warn her when and where she has to go, or tell her 
which facilities are available in the airport lounge. Yet not all 
devices at the airport may be ready to help a PDA, e.g. because of 
security restrictions, because they are proprietary and reserved for 
paying customers, or because they simply do not care about personal 
wishes. Moreover, devices may be ready to share certain types of 
services but not others, e.g. telling users when the flight is 
scheduled to depart, but not how many passengers will be on it. As 
another example, devices may not only be uncooperative, but 
malevolent, in the sense that they try to manipulate other devices in 
a way detrimental to their user. Such devices may be programmed, e.g. 
by fraudsters, spies, or terrorists.

	There exists an extensive literature on the evolution of cooperation 
between initially ``selfish" agents, inspired by the seminal work of 
Axelrod \cite{Axelrod1984} that compared different strategies for 
playing a repeated ``Prisoners' Dilemma" game. However, this game 
does not seem directly applicable to information exchanging devices. 
Moreover, the chief result, while sensible, may seem trivial: the 
most effective strategy to achieve robust cooperation appears to be 
\emph{tit for tat}, i.e. cooperate with agents that reciprocate the 
cooperation, stop cooperating with those that do not. More recent, 
tag-based models (e.g. \cite{RioloEtAl2001, HalesEdmonds2003} start 
from a simpler situation than the Prisoners' Dilemma, in which one 
agent ``donates" a service to another one, at a small cost to the 
donor but a larger benefit to the recipient. The main idea is that 
agents are identified by ``tags", and that they cooperate with those 
agents whose tags are similar to their own. The rationale is that 
agents with the same type of tag belong to the same group, ``family" 
or ``culture", following the same rules, so that they can be trusted 
to reciprocate.

	For artifacts, a tag may include such markers as brand, 
model, and protocols understood. This would show that a device is 
capable and willing to lend particular services to another one, thus 
obviating the need for a repeated, ``tit-for-tat-like" interaction 
probing the willingness to reciprocate. Yet extensible environments 
should allow the addition of very dissimilar devices, made by 
different companies using different standards and functionalities. 
Therefore, we propose a different approach, combining some advantages 
of tags and tit-for-tat strategies.

	Consider a game with the following moves: an agent makes a 
request and the other agent either cooperates (donates) or 
``defects". Agents learn from these interactions in the following 
manner: if the result is positive (cooperation), the agent will get 
more ``trust" in the other agent's cooperativeness. Thus, the 
probability increases that it will make further requests to that 
agent in the future, or react positively to the other's requests. 
Vice-versa, a negative result will lead to more ``distrust" and a 
reduced probability to make or accept requests to/from this agent. 
Still, to recognise this agent, it has to take its clue from the tag, 
which is usually not unique to that agent. This means that a later 
interaction may be initiated with a different agent that carries a 
similar tag, but that is not necessarily willing to cooperate to the 
same extent. We may assume that if the first few interactions with 
agents having similar tags all generate positive (negative) results, 
the agent will develop a default propensity to react positively 
(negatively) always to agents characterised by that type of markers.

	We expect that in this way the initially undirected 
interactions will produce a differentiation in clusters of similarly 
marked agents that cooperate with each other (e.g. all devices 
belonging to the same user or organization), but that are reluctant 
to interact with members of other groups (e.g. devices belonging to 
rival organizations). The tags and their association thus develop the 
function of a mediator \cite{Heylighen2003} that increases the 
probability of positive interactions by creating a division between 
``friends" (in-group) and ``strangers" or ``foes" (out-group). Note, 
however, that there is no assumption that an agent only cooperates 
with agents bearing the same tag as itself: by default it cooperates 
with anyone having a tag similar to the one of agents that were 
cooperative in the past. This means that there can be groups with 
which everyone cooperates (e.g. ``public" devices), but also that 
specific types of ``symbiosis" can develop in which one group 
systematically seeks out members of a different group to cooperate 
with because of their complementary capabilities. This brings us to 
the more complex issue of the division of labour.

\section{Learning to coordinate}

	After having ascertained that our devices can communicate and 
cooperate, we still need to make sure that the functions they perform 
satisfy the user. This desired functionality can be viewed as a 
complex of tasks that need to be executed. The tasks are mutually 
dependent in the sense that a certain task (e.g. locating a file) has 
to be accomplished before subsequent tasks (e.g. downloading and 
playing the file) can be initiated. Each agent can either execute a 
task itself, or delegate it to another agent. Initially, we may 
assume that all agents that have a certain functionality built in 
(e.g. playing a sound file) are equally competent at performing that 
type of task. However, in practice the satisfaction of the user can 
vary. For example, a recording is likely to be played with a higher 
sound quality by a music installation than by a PDA or television. By 
default, devices can use certain preprogramed rules-of-thumb to 
decide who takes precedence (e.g. newer or more specialized devices 
are preferred to older, less specialized ones). Yet in an open 
environment there is no guarantee that such simple heuristics will 
produce the best result. Again, we may tackle this problem through 
individual learning coupled with collective self-organization.

	Assume that the user regularly expresses his/her overall 
satisfaction with the ambient intelligence environment (e.g. 
explicitly by clicking on a scale from one to ten, or implicitly by 
facial or physiological cues that express happiness/unhappiness). 
This score can be used as a feedback signal to the network of 
devices, allowing it to reinforce the more successful rules, while 
weakening the less effective ones. We will assume that the agent who 
delegated a task will increase its trust in the competence of the 
agent that performed that task, and thus increase its probability to 
delegate a similar task to the same agent in the future. Otherwise, 
it will reduce its trust. As demonstrated by the simulation of Gaines 
\cite{Gaines1994}, this assumption is sufficient to evolve a 
self-reinforcing division of labour where tasks are delegated to the 
most ``expert" agents.

	However, when the tasks are mutually dependent, selecting the 
right specialist to carry out a task is not sufficient: First the 
preparatory tasks have to be done by the right agents, in the right 
order. When the agents do not know a priori what the right order is, 
they can randomly attempt to execute or delegate a task, and, if this 
fails, pick out another task. Eventually they will find a task they 
can execute, either because it requires no preparation, or because a 
preparatory task has already been accomplished by another agent. Each 
completed task enables the accomplishment of a series of directly 
dependent tasks. In this way the overall problem will eventually be 
solved. In each problem cycle, agents will learn better when to take 
on which task by themselves, or when to delegate it to a specific 
other agent.

We expect that this learned organisation will eventually stabilise 
into a system of efficient, coordinated actions, adapted to the task 
structure. When new devices are added to the system, system and 
device should mutually adapt, producing a new organization. While no 
single agent knows how to tackle the entire problem, the knowledge 
has been ``distributed" across the system. The ``tags" that identify 
agents, and the learned associations between a tag and the competence 
for a particular task, play the role of a mediator 
\cite{Heylighen2003}, delegating tasks to the right agents and 
coordinating their interactions so that the problem is tackled as 
efficiently as possible.

\section{Conclusions}

We cannot keep on adding functions to personal computers. They serve 
as text editors, game consoles, televisions, home cinemas, radios, 
agendas, music players, gateway to the Internet, etc. Such general 
devices will never produce the same quality as specialized 
appliances. Our PCs are like ducks: they can swim, but not as well as 
fish; fly, but not as well as hawks; and walk, but not as well as 
cats. Rather than integrate so many functions in a single device, it 
seems preferable to entrust them to an ever expanding network of 
specialized devices that is kept coordinated through an ongoing 
process of self-organization. We have described a number of general 
requirements and approaches that may enable our artifacts to learn 
the most effective way of cooperation.

In our overall scenario, we have assumed that standard functions and 
interaction rules are preprogrammed by a global community to handle 
the most common, default situations, but that the system is moreover 
ready to extend its own capabilities, adapting to newly encountered 
tasks, situations, or devices. This ability to adapt should be 
already present in the interaction rules. The adaptation may be 
achieved through the self-organization of the system of agents, using 
recurrent, ``game-like" interactions, in which the agents learn what 
messages mean and who they can trust to perform which task. Most of 
this can happen outside of, or in parallel with, their normal 
``work", using idle processing power to explore many different 
communication and collaboration configurations. Thus, we can imagine 
that our future, intelligent devices, like young animals or children, 
will learn to become more skilful by exploring, ``playing games" with 
each other, and practising uncommon routines, so as to be prepared 
whenever the need for this kind of coordinated action appears.

\section{Acknowledgements}

We thank Peter McBurney for useful comments. C. G. was supported in 
part by CONACyT of Mexico.

\bibliographystyle{ICCS}
\bibliography{carlos,sos}

\end{document}